\documentstyle[emulateapj,psfig]{article}

\def\etal{{\it et al. }}

\def\simlt{\mathrel{\spose{\lower 3pt\hbox{$\mathchar"218$}}
     \raise 2.0pt\hbox{$\mathchar"13C$}}}
\def\simgt{\mathrel{\spose{\lower 3pt\hbox{$\mathchar"218$}}
'     \raise 2.0pt\hbox{$\mathchar"13E$}}}
\def\gsim{ \lower .75ex \hbox{$\sim$} \llap{\raise .27ex \hbox{$>$}} }
\def\lsim{ \lower .75ex \hbox{$\sim$} \llap{\raise .27ex \hbox{$<$}} }
\newcommand{\kpch}{\, h^{-1}{\rm kpc}}    

\newcommand{\Msunh}{h^{-1}M_{\odot}}
\newcommand{\Mpch}{\, h^{-1}{\rm Mpc}} 
\long\def\***#1{{\scshape ***#1***}}
\def\arraystretch{1.25}


\begin{document}

\lefthead{GOVERNATO ET AL.} 
\righthead{DESCENDENTS OF LYMAN BREAK GALAXIES}

\title{The descendents of Lyman break galaxies in galaxy clusters:  \\
 spatial  distribution and orbital properties} 
  \author{F.~Governato\altaffilmark{1},
  S.~Ghigna\altaffilmark{2,4},B.Moore\altaffilmark{2}, T.~Quinn\altaffilmark{3}, J.~Stadel\altaffilmark{3} \& G.~Lake\altaffilmark{3}}

\altaffiltext{1}{Osservatorio Astronomico di Brera, Merate, Italy}

\altaffiltext{2}{Physics Department, University of Durham, Durham City, UK}

\altaffiltext{3}{Astronomy Department, University of Washington, Seattle, USA}

\altaffiltext{4}{Dipartimento di Fisica, Universit\'a di Milano-Bicocca, 
Milan, Italy}

\begin{abstract}

We combine semi--analytical methods with a ultra-high resolution
simulation of a cluster (of mass $2.3 \times
10^{14}\Msunh$, and 4 $\times$ 10$^6$ particles within its
virial radius) formed in a standard CDM universe to study the spatial
distribution and orbital properties of the present--day descendents of
Lyman Break galaxies (LBG). At redshift 3 we find a total of 12 halos
containing at least one Lyman Break galaxy in the region that will
later collapse to form the cluster itself.  At the present time only
five of these halos survive as separate entities inside the virial
radius, having been stripped of most of their dark matter. Their
circular velocities are in the range 200 -- 550 km/sec.  Seven halos
merged together to form the central object at the very center of the
cluster.  Using semi-analytical modeling of galaxy evolution we show
that descendents of halos containing Lyman Break galaxies now host
giant elliptical galaxies such as those typically found in rich galaxy
clusters. All galaxy orbits are very radial, with a pericenter to
apocenter ratio of about 1:5.  The orbital eccentricities of LBG
descendents are statistically indistinguishable from those of the
average galaxy population inside the cluster, suggesting that the
orbits of these galaxies are not significantly affected by dynamical
friction decay after the formation of the cluster's main body.  In
this cluster, possibly due to its early formation time, the
descendents of Lyman break galaxies are contained within the central
60\% of the cluster virial radius and have an orbital velocity
dispersion lower than the global galaxy population, originating a mild
luminosity segregation for the brightest cluster members.  Mass
estimates based only on LBG descendents (especially including the
central cD) reflect this bias in space and velocity and underestimate
the total mass of this well virialized cluster by up to a factor of
two compared to estimates using at least 20 cluster members.

\end{abstract}


\section{Introduction}

The implementation of a simple color selection technique to select
efficiently galaxies at redshift larger than 2.5 (Steidel \etal 1996,
Madau \etal 1996, Steidel \etal 1999, Fontana \etal 1999 ) revealed a
population of blue, actively star forming galaxies at high redshift.
Galaxies as bright as those observed are likely hosted inside the most
massive halos at high z (Bagla 1998, Baugh \etal 1998, Katz, 
Hernquist \& Weinberg 1999, Coles \etal 1998, but
see Somerville, Primack \& Faber 1998 and Kolatt \etal 1999 for a
slightly different view). These halos are more clustered (a bias
factor of the order of 2-5) compared to the general distribution,
providing strong support (Adelberger \etal 1998, Giavalisco \etal
1998) to models of biased galaxy formation (Davis \etal 1985). Under
the effect of gravitational instability these large halos will merge
together and form the massive clusters we see today (Governato \etal 1998).
Semi--analytical models (Baugh \etal 1998) further suggested that the
present day descendents of LBG in protoclusters would be
preferentially giant ellipticals with an old red population of stars.

Bright, red cluster members reside preferentially at the center of
clusters and often have been found to have a lower orbital velocity
dispersion (Chincarini \& Rood 1977, Mellier \etal 1988, Biviano
\etal 1992, Whitmore \etal 1993, Biviano \etal 1996, Carlberg \etal
1997) than the global cluster population.  Recent results with full
redshift information for a large sample of clusters (Adami, Biviano \&
Mazure 1998, Ramirez, de Souza \& Schade 2000) and photometric
observations of the Coma cluster (Kashikawa 1998) give support to
these claims.  Adami \etal, based on a simple theoretical modeling,
suggest that orbits of the brightest galaxies have to be circular to
explain the decrease in velocity dispersion and at the same time be
consistent with the hypothesis of dynamical equilibrium at the cluster
center.

\begin{figure*} 
\centerline{ \hbox{
\psfig{figure=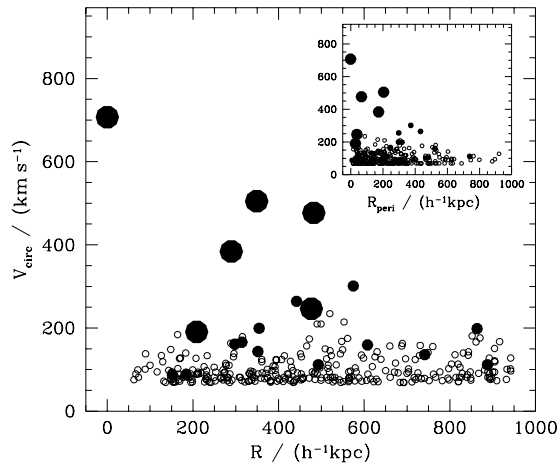,height=8cm,width=8cm,angle=0} } }
\caption{ Circular velocity V$_c$ (see text) vs.  distance from the
cluster center at $z=0.1$. Large filled dots are descendents of LBGs,
small filled dots are the 20 brightest cluster members. Small empty
circles show the whole cluster population with V$_c$ $>$ 70 Km/sec.  
Inner panel: V$_c$ vs.  orbital
pericenter. Symbols are as in the main panel.}
\label{fig1} 
\end{figure*}

Indeed theoretical prejudice would expect to find galaxies formed in
massive halos at high redshift to reside preferentially in the central
region of clusters.  Moore \etal (1998) and White \& Springel (1999)
showed that, in CDM cosmologies, matter already in virialized objects
at high redshift makes a large fraction of the mass within the
central regions of present day clusters.  Dynamical friction, if
acting efficiently on a long enough time scale could further segregate
massive halos at the center of clusters (but the effect is likely to
be small; see Ghigna \etal 1998 and Colpi, Mayer \& Governato 1999,
hereafter CMG99). In recent years, numerical and analytical studies of
galaxy clusters have rapidly increased in resolution and detail (e.g.
Katz \& White 1993, Carlberg 1994, Frenk \etal 1996, Fusco--Femiano \&
Menci 1998, Tormen, Diaferio \& Syer 1998, Klypin \etal 1999). Frenk
\etal (1996) found mild spatial segregation of the most massive
galaxies, they also included gas dynamics and a simple description of
star formation processes.

In this work we use a unified approach that couples a state of the art
numerical simulation of a galaxy cluster with a detailed,
semi--analytical description of galaxy formation inside individual
dark matter halos. This method will allow us to study the spatial and
orbital distribution of galaxies in a moderately rich cluster with
unprecedented detail and to investigate the relation between LBGs at
high redshift and present day cluster galaxies.

\section{Coupling N-body simulations and semi--analytical models}

\subsection {The Cluster simulation}   

We used a very high resolution N-body (i.e. collisionless) simulation
of a galaxy cluster (slightly more massive than Virgo, $2.3 \times
10^{14} \Msunh$ within the virial radius, defined as the radius where
$\rho(r<R) \sim 200 \rho_{crit}$) formed in a cluster normalized SCDM
cosmology.  This cluster contains over 4 $\times$ 10$^{6}$ particles
within the virial radius (it is described in full detail in Ghigna
\etal 1998, G98, Ghigna \etal 1999, G99, and Lewis {\it \etal} 1999).
The effective spatial resolution is of the order of $1.0\kpch$, and we
are able to resolve substructure halos with circular velocities V$_c$
down to 50 km/sec and with pericenters larger than $50\kpch$, a
significant improvement compared with all previous works (here V$_c$
is defined as $\sqrt{ G M(<r)/r }$)).

The cluster forms through major mergers at redshift about 0.5 (defined
as when its main progenitor has roughly 50\% of the final cluster
mass), accreting additional mass and galaxies at later times. It is
well virialized and close to dynamical equilibrium by the present time
(see G98 fig.1).  In this high resolution simulation, numerical
overmerging (e.g., Moore, Katz \& Lake 1996) is likely to be almost
negligible, especially for the most massive halos. We can follow the
evolution of thousands of halos as they participate in the build up of
the cluster and subsequently orbit inside it.  Even if severely
stripped by the cluster tidal field, virtually all halos maintain
their identity once inside the cluster, and only a few get destroyed
by the tidal field or decay by dynamical friction at its center (see
G99 for a full discussion).

Therefore, within this simulation is possible to follow the
descendents of all halos and specifically those associated with LBG
(see next subsection) through subsequent outputs to the present time.
We first located dark matter halos inside the clusters with the
algorithm SKID (see G98 and G99 for details, Springel 1999 for an
alternative method) at the final time of the simulation and traced
them back to dark matter halos at z = 3. High z halos were identified
with "Friends--of--Friends" (FOF, Davis \etal 1985), using a linking length 
$= 0.2$ the initial grid spacing,  as FOF gives more robust
masses for halos outside larger virialized structures. At $z=3$ the
region containing the cluster has yet to collapse, but hundreds of
smaller halos have already formed within a complex network of
filaments.  For each halo we measured mass and circular velocity at
their virial radius at high z and at their individual tidal radii as
imposed by the cluster potential at the present time (again see G98
for details).

\subsection{Semi-analytical galaxy formation}

The growth of dark matter halos can be followed both with N--body
simulations and the extended Press \& Schechter approach (or PS, see
Press \& Schechter 1974, Bower 1991, Bond \etal 1991). Within the
semi--analytical approach a simple set of equations then describes the
cooling of gas inside the dark matter halos and the subsequent star
formation history, predicting size, luminosities (including the
effects of dust), colors and the morphology of galaxies formed inside
these halos (Kauffmann, White \& Guiderdoni 1993, Cole \etal 1994,
Cole \etal 2000). We used the approach first outlined in Cole \etal
1994 and then further developed in Baugh \etal 1998 and Cole \etal
2000 This approach is remarkably successful in predicting the main
properties of high redshift and local galaxies with a minimum set of
constrained parameters (Baugh \etal 1998, see also Somerville \&
Primack 2000).  However, the semi-analytical method based on the PS
alone cannot recover the full 3D distribution of galaxies, making the
full N-body simulations necessary.

For each halo identified at redshift 3 in the N--body simulation and
for the whole cluster at the present time we used the semi-analytical
approach (and so a PS merging history) to determine their galaxy
content.  At the present time the halos present inside the cluster are
paired with semi-analytical cluster galaxies based on their
circular velocity. Statistically this is equivalent to using merger
trees obtained directly from N-body simulations (see Governato \etal
1998,  Benson et al. 1999).  As a test we looked at
a set of different merger tree histories for a few clusters of the
same mass as the one in our simulation to verify that the scatter
introduced by our approach on the average properties as a function of
circular velocities of the galaxy population was negligible.  In fact,
the galaxies produced with the semi-analytical approach show a rather
tight luminosity-circular velocity relation, independently of the
details of their merging history or of that of the  parent cluster.  This
simple approach is then quite adequate for our  purpose of
broadly defining the types of galaxies hosted both in massive
halos at high z and inside their cluster descendents at the present
time as a function of their mass (see Kauffmann \etal 1999, Springel
1999 for an approach based on the full merger trees obtained from N--body
simulations).

Our procedure gives the properties of the galaxies hosted inside each
given halo complete with full dynamical information (position and
velocity inside the cluster). Once galaxies were placed inside dark
matter halos, we selected at redshift of 3 those that, applying the
same criteria, would have been selected as LBGs (Steidel \etal (1996).
At the final time we then compare the properties of their descendents
vs. those of the 20 brightest cluster members (comparable to the
number of redshift usually measured for a single real cluster) and the
whole cluster galaxy population.

\section{Results: The descendents of Lyman Break Galaxies}

At a redshift of 3 there are 12 halos with mass above 10$^{12}$
M$_{\odot}$ (the biggest object in the region that will later form the
cluster has a mass of $3.2 \times 10^{12} \Msunh$).
The semi-analytical approach predicts that each of these halos hosts at
least one LBG galaxy, sometimes two. Indeed the N-body simulation
already shows significant substructure inside them.  There is some
intrinsic scatter from one semi-analytical realization to another,
depending on the details of the merging histories of individual
halos. Sometimes smaller halos (on average less than one per
realization) host LBGs, perhaps ``observed'' while they were at their
maximum luminosity. This does not change our results significantly.

Halos containing LBGs are aligned along filaments and are rapidly
flowing along them to form massive groups at z $\sim$ 1.5--0.75 and
then merge to form the main progenitor of the cluster by z$=$0.5.  A
large fraction (7 out of 12) merge together to form the central core
of the cluster; 90\% of the mass contained in their central part
(defined as particles within the central $10\kpch$) and likely
tracers of their stellar component ends up in the inner $125\kpch$ 
of the cluster.  Their barionic cores (not present in our
simulation that includes only the dark, collisionless component) would
then most likely merge together to form the central cD, as the decay
time for any remnant of significant mass with orbital apocenters less
than $100\kpch$ from the cluster center is much shorter than the
Hubble time.  The five surviving halos have been tidally stripped and
orbit in the central part of the cluster. According to the
semi--analytical approach the descendents of LBGs are the most
luminous ellipticals in the cluster at the present day.  This result
is independent of the details of the semi--analytical model used. In
the approach used e.g. in Kolatt \etal (1999) a large number of Lyman
Break Galaxies are small starbursting galaxies. These strong episodes
of star formation originate from fly-bys between satellites inside
more massive halos.  In principle, these satellites could be stripped
away from their parent halos and show a different spatial bias, making
our conclusions dependent on the analytical modelling.

However, our simulation shows that none of the satellites of
the massive halos at z = 3 survive as distinct entities by the present
time, having merged with their hosts before the formation of the main
body of the cluster.

\begin{figure*} 
\centerline{\hbox{
\psfig{figure=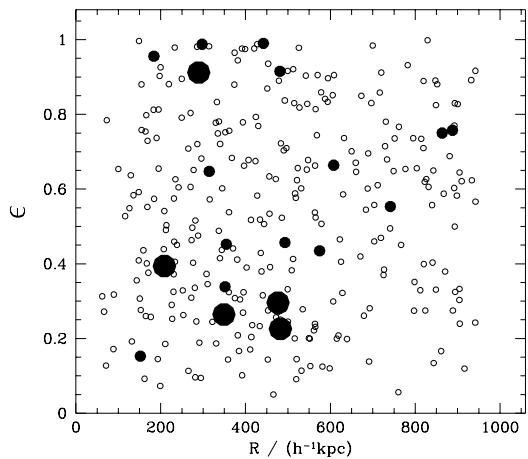,height=8cm,width=8cm,angle=0} }}
\caption{ Orbital circularity J/J$_c$ (see text) vs. radial distance
from the cluster center at $z=0.1$. Large filled dots are
 galaxies descendents of LBG, small filled dots are the 20 brightest
cluster members. Small open circles the whole cluster population with
V$_c$ $>$ 70 Km/sec.  
}
\label{fig2} 
\end{figure*}

\subsection{Orbits and luminosity segregation}

At $z=0.1$ all LBG descendents can be found within the inner
$0.6\Mpch$, i.e 60\% of the virial radius of the cluster. They are
more concentrated than the average cluster population (see
Fig.~1). This is more evident in the distribution of the pericentric
distances (inset of Fig.~1) and is true for apocenters as well.  To
strengthen the significance of the signal, we have verified that this
holds true at a nearby epoch ($z=0$). Using Wilcoxon
test, we estimate that the probability of this spatial segregation happening by
chance is less than 2\%.  Also considering that seven LBGs contributed
to form the central galaxy, the mass contributed to the cluster by LBG
descendents is more centrally concentrated compared with the global
cluster population.  As halos with large circular velocities are
associated with galaxies of higher luminosities than the average
galaxy cluster population, this causes a mild luminosity segregation.
It is likely that the early formation time of this cluster and its
following quiet merging history (it forms slightly earlier than
average for its mass in a SCDM cosmology; G98) contributed to this,
as recent infall was not substantial enough to accrete massive halos
at the outskirts of the cluster.

We then measured the orbital parameters for all galaxies inside our cluster
(see Fig.~2). The orbital circularity $\epsilon$ is defined as the
ratio of J, the orbital angular momentum, to J$_c$, the angular
momentum of a circular orbit with the same energy. (Lacey \& Cole
1994, Tormen 1997, G98). Here the orbital energy is defined assuming
spherical symmetry for the cluster mass distribution and the most
bound particles for its center.  There is no obvious difference in the
distribution of circularities between descendents of LBG, the twenty
brightest objects in the cluster and the rest of the galaxy
population. The formal average values of $\epsilon$ for the three cases are
$0.42$, $0.59$ and $0.54$, respectively, with quite similar
dispersions around the mean value ($\sim 0.3$). The dark matter
background has similar orbital properties (G98). Results at $z=0.1$ and
$0$ are similar.

 This finding confirms results obtained with analytical and numerical
 models (van den Bosch \etal 1998, CMG99) that dynamical friction is not
 efficient at circularizing orbits of even the most massive and old
 galaxies inside clusters.  We used the theory of linear response as
 described in CMG (which agrees extremely well with N-body
 experiments) to measure the orbital decay predicted for a group sized
 halo entering the cluster environment at $z=1$ (i.e. the formation time
 of the main progenitor of the cluster itself).  Once the effect of
 tidal stripping are included (see again CMG99) decay times are of the
 order of several times the Hubble time, and both pericenters and
 apocenters have changed only by a few percent by the present time.
 The luminosity segregation putatively observed in real galaxy clusters
 would then be an imprint of their hierarchical build up rather than
 the effect of subsequent strong dynamical evolution.  This
 orbital segregation should be present (Springel, 2000 in preparation)
 or could even be larger in clusters formed in a open or flat
 cosmology, where clusters would form typically at higher redshift and
 where the accretion at late times slows down considerably.

Our simulation allows us for the first time to test the dynamical mass
estimate based on a complete sample of substructure halos.  We
estimated the virial mass of the cluster from the galaxies' projected
velocity dispersions, using the classic estimator (Heisler, Bachall \&
Tremaine, 1985):

$ M_{VT} = ( 3 \pi N / 2G ) {\sum{_i}v^2_{p,i}\over{\sum_{i<j} R_{ij}^{-1}}}  $

where v$_{p,i}$ is the line of sight velocity and R$_{ij}$ the
projected separation of a given galaxy pair. This estimator is useful
for its simplicity, even if it overestimates the mass inside  the
virial radius by about 40\% (see also Girardi \etal 1998 and
references therein).  We do not include galaxies in halos outside the
virial radius, so that our sample is free of nearby back/foreground
interlopers.  Our results confirm previous results 
(Frenk \etal 1996, Tormen 1997) 
that the use of only the few brightest galaxies as mass
estimators results in an underestimate of the cluster mass compared to
using the whole galaxy sample, by up to a factor of 2 if the brightest
galaxy is included (it has a very small velocity compared to the
cluster galaxies as a whole). Additional scatter ($\sim$30\%) is added
when considering individual axial projections.  Contrary to previous
suggestions, this bias is not due to the most massive galaxies being
on more circular orbits, but the fact that these galaxies sample only
the central part of the cluster mass distribution and therefore have a
lower velocity dispersions, as the peak in V$_c$ {\it for the cluster
as a whole} is reached only at about $0.5\Mpch$, i.e. close to
the apocenters of their orbits.  Even excluding contamination from
background and foreground objects  at least 20 galaxies are needed
to correctly sample the cluster potential and obtain a reliable mass
measurement. Likely, even more redshifts would be needed in case the
cluster had significant nearby structures (filaments or rich groups)
in the near fore/background.

\section{Discussion}

Using the high resolution simulation of a galaxy cluster coupled with
  semi--analytical methods of galaxy formation we identify at redshift
  of 3 a dozen halos hosting at least one Lyman Break Galaxy. At the
  present time descendents of LBGs can be identified with the central
  cD galaxy and galaxies hosted in substructure halos with V$_c$ in
  the range 200 to 550 km/sec.  All 12 LBG descendents 
 end up within the the inner
  $\sim 0.5\Mpch$ (or 60\% of the cluster virial radius); 7 merged
  together to form the central galaxy. These
  descendents are the most bright elliptical galaxies in the cluster.
  These results are largely independent from the details of the
  semi-analytical method used. We confirm previous findings obtained
  with simulations of lower resolution (e.g Frenk \etal 1996) that the
  most massive galaxies are likely to be centrally segregated and have
  lower orbital velocity dispersions when compared to the global
  cluster galaxy population. However, this effect is small, and harder
  to detect when only limited information (redshifts and positions
  projected  on the sky plane) is available.

 Spatial and velocity segregation for bright cluster members has long
 been observed in Coma (Mellier \etal 1988) and in larger samples of
 nearby clusters (Biviano \etal 1992, 1996), but the observational
 picture has been somewhat complicated by the small number of redshift
 available per cluster and by the fact that they have usually been
 collected only within the central part of the clusters
 themselves. Clearly a larger sample of observed and simulated
 clusters is needed to allow a more quantitative comparison between
 observations and theoretical predictions.  We expect the segregation
 of bright ellipticals to be larger in well virialized clusters and in
 cosmologies were recent infall is small (e.g. open or flat CDM
 cosmologies).

Galaxies in our simulated cluster move on quite eccentric orbits, due
to the almost radial infall typical of hierarchical clustering. There
is no significant difference in the orbital eccentricity of different
galaxy populations and the dark matter background. Also, orbits do not
change in shape significantly over time (dynamical friction does not change
the orbital eccentricity  as shown also in CMG99).

 Our analysis shows that to measure the virial mass of the cluster is
crucial that a significant number of galaxies is used in order to
correctly sample the cluster density profile. A sample, restricted to
the most bright cluster members is likely to be biased and
underestimate both the cluster total mass and velocity dispersion.  In
this  simulated cluster, about twenty member galaxies sampling the mass
distribution out to the virial radius are required for a correct
estimate of the cluster total mass. This number could well be higher
for a cluster far from virial equilibrium or with significant
structures nearby. As our analysis shows, virial mass estimates 
suffer from an additional scatter of about 30\%, due to velocity
anisotropies along the cluster projection. This source of scatter
cannot easily be removed increasing the number of galaxies.

As clusters likely   formed only a few Gyrs ago,  dynamical effects
like energy equipartion or dynamical friction are very unlikely to
have played any significant role in originating the mass/velocity
segregation, especially considering that only a small part of the cluster
mass is attached to individual galaxies ($\lsim 15\%$, see G98). 
If confirmed by a larger sample of real and simulated
clusters, the observed segregation of their more massive galaxies
would rather be the signature of their hierarchical build--up.

\section{Acknowledgements} 

We thank Carlton Baugh and Andrew Benson for providing us the
semianalytical galaxy formation models and acknowledge useful
discussions with Frank van den Bosch, Rychard Bouwens and Cedric
Lacey.  This work was partially supported by the EU Network for Galaxy
Formation and Evolution, the NASA/ESS programme, and PPARC. SG is a
Marie Curie fellow, while BM is a Royal Society Research Fellow.
Simulations were run at the Edinburgh supercomputing center and NCSA.

\end{document}